\documentclass[letterpaper, 10 pt, conference]{ieeeconf}  
\usepackage{amsmath}
\usepackage{amssymb}
\usepackage{graphicx}
\usepackage[colorinlistoftodos]{todonotes}
\usepackage{url}
\usepackage[colorlinks=true, allcolors=blue]{hyperref}
\usepackage{booktabs}
\usepackage{siunitx}
\usepackage{amsfonts}
\usepackage[noadjust]{cite}
\usepackage{algorithm}
\usepackage{algorithmicx}

\renewcommand{\baselinestretch}{1.25}

\newcommand{\Rn}{\mathbb{R}^n}
\newcommand{\Rm}{\mathbb{R}^m}
\newcommand{\Xc}{\mathcal{X}}
\newcommand{\Uc}{\mathcal{U}}
\newcommand{\Sc}{\mathcal{S}}
\newcommand{\dxy}{
\begin{bmatrix}
 \dot{x} \\ \dot{y}
 \end{bmatrix}
}
\newcommand{\dxysq}{
\begin{bmatrix}
 \dot{x}^2 \\ \dot{y}^2
 \end{bmatrix}
}
\newcommand{\dxycube}{
\begin{bmatrix}
 \dot{x}^3 \\ \dot{y}^3
 \end{bmatrix}
}
\newcommand{\dxyquad}{
\begin{bmatrix}
 \dot{x}^4 \\ \dot{y}^4
 \end{bmatrix}
}
\newcommand{\diagdxy}{
\begin{bmatrix}
 \dot{x} & 0 \\
 0 & \dot{y}
 \end{bmatrix}
}

\newcommand{\ddxy}{
\begin{bmatrix}
 \ddot{x} \\ \ddot{y}
 \end{bmatrix}
}
\newcommand{\ddxysq}{
\begin{bmatrix}
 \ddot{x}^2 \\ \ddot{y}^2
 \end{bmatrix}
}

\newcommand{\dddxy}{
\begin{bmatrix}
 \dddot{x} \\ \dddot{y}
 \end{bmatrix}
}

\DeclareMathOperator*{\argmin}{arg\,min}

\IEEEoverridecommandlockouts                              
\title{\LARGE \bf
Barrier Functions in Cascaded Controller: Safe Quadrotor Control
}

\author{Mouhyemen Khan, Munzir Zafar, Abhijit Chatterjee
\thanks{Mouhyemen Khan, Munzir Zafar, and Abhijit Chatterjee are with the School of Electrical and Computer Engineering, Georgia Institute of Technology, Atlanta, GA 30332, USA: {\tt\small \{mouhyemen.khan, mzafar7, abhijit.chatterjee\}@gatech.edu }
}
}

\begin{document}

\maketitle
\thispagestyle{empty}
\pagestyle{empty}

\begin{abstract}
Safe control for inherently unstable systems such as quadrotors is crucial. Imposing multiple dynamic constraints simultaneously on the states for safety regulation can be a challenging problem. In this paper, we propose a quadratic programming (QP) based approach on a cascaded control architecture for quadrotors to enforce safety. Safety regions are constructed using control barrier functions (CBF) while explicitly considering the nonlinear underactuated dynamics of the quadrotor. The safety regions constructed using CBFs establish a non-conservative forward invariant safe region for quadrotor navigation. Barriers imposed across the cascaded architecture allow independent safety regulation in the quadrotor's altitude and lateral domains. Despite barriers appearing in a cascaded fashion, we show preservation of safety for quadrotor motion in SE(3). We demonstrate the feasibility of our method on a quadrotor in simulation with static and dynamic constraints enforced on the position and velocity spaces simultaneously.
\end{abstract}

\section{Introduction}\label{sec:intro}
Safety is a critical component for today's autonomous aerial systems \cite{safe1}, \cite{safe2}, \cite{safe3}. Of particular interest are quadrotors due to their application in surveillance, agriculture, and acrobatic performances, see \cite{quadapp1}, \cite{quadapp2}, \cite{quad_dance}. Thus accentuating the need for safety as an imperative requirement during flight operation. Control Barrier Functions (CBFs) have proven to be an effective strategy for guaranteeing safety in several applications \cite{cbf},\cite{biped}, including quadrotors \cite{li-swarm-mob}. The focus of this paper is to rectify the quadrotor's nominal trajectory using a cascaded controller to ensure safety in 3D position and velocity spaces. We achieve this by independently imposing barriers across the cascaded hierarchy.


The underactuated and intrinsically unstable nature of quadrotor makes it challenging to generate safe trajectories \cite{quad_traj}. CBFs, first implemented in adaptive cruise control \cite{cbf} formed as an online quadratic program (QP), permit dynamically feasible constraints and ensure forward invariance. CBFs were used in collision avoidance for swarms of mobile ground robots \cite{li-swarm-mob} and quadrotors \cite{li-swarm-quad}. They restricted their safety constraints to the position space only. CBFs were also used to learn quadrotor dynamics in the presence of wind disturbances \cite{li-safe}. This work uses a differential flatness model and a single CBF in its controller scheme with a focus on learning unmodeled dynamics. The works in \cite{2dquad} and \cite{3dquad} use sequential QP based methods augmenting the CBF with a Control Lyapunov Function (CLF) for obstacle avoidance resulting in a CLF-CBF-QP controller. They propose a sequential optimization scheme where virtual thrust is first computed using position level QP satisfying a CLF constraint, while CBF constraints are incorporated in the lower level QP to generate the control inputs. Their controller, therefore, results in a sequential CLF-CBF-QP formulation. The work in \cite{2dquad} and \cite{3dquad} impose safety explicitly in $SE(2)$ and $SE(3)$ respectively. In order to find a feasible solution for QP, input bounds are not considered in \cite{3dquad} (although quadrotors have input bounds, e.g. no reverse thrust).




\textbf{Key contributions:} In constrast to prior research, our contributions in this paper are threefold:
\vspace{-0.25cm}
\begin{itemize}
	\item Prior work has not merged the forward invariance of CBFs at \textit{every level of the hierarchy} in cascaded controller to ensure safety for quadrotors. The cascaded control scheme is a popular architecture and imposing safety on such a controller is yet to be formulated. We present derivations for enforcing constraints across the hierarchy by considering the complete 3D quadrotor dynamics evolving in tangent bundle to $SE(3)$.
    \item CBFs are employed in a nonlinear cascaded controller with constraints \textit{explicitly} imposed on the position and velocity spaces. For safety-critical tracking of complex trajectories, imposing safety on velocity along with position is critical. We have also empirically verified in simulation that our QP formulation with actuator bounds found feasible solution. To the best of our knowledge, this is the first time constraints on position and velocity spaces are explicitly formulated for safe 3D quadrotor control along with actuator bounds.
	\item Safety constraints are handled in the altitude and lateral domains of the quadrotor \textit{independently}. We place separate CBF constraints on the outer loop and inner loop of the controller. This has two advantages: (i) decoupling the safety constraint results in a richer superlevel safe set and (ii) this allows independent regulation of quadrotor flight in altitude and lateral domains.
\end{itemize}

The rest of the paper is organized as follows. Section \ref{sec:prelim} introduces preliminaries on barrier functions and quadrotor dynamics. Cascaded controller with augmented QP design is discussed in Section \ref{sec:controller} while safety barrier formulations across the hierarchy are derived in Section \ref{sec:safety-regions}. Simulation results are provided in Section \ref{sec:simulation}, followed by conclusions in Section \ref{sec:conclusion}.


\section{Preliminaries: Barrier Functions and Quadrotor Dynamics}\label{sec:prelim}
This section introduces CBFs along with its extension called Exponential CBFs (ECBFs) and dynamics of a quadrotor in 3D. These topics are well studied, hence for a more detailed discussion on CBFs, ECBFs, and quadrotor dynamics, we refer the reader to \cite{cbf}, \cite{ecbf}, and \cite{corkekumar} respectively.

\subsection{Control Barrier Functions}
Consider a general control affine dynamical system,
\vspace{-0.2cm}
\begin{align}\label{affine}
\dot{x} = f(x) + g(x)u,	\ \ \ \		x(t_0) = x_0,
\end{align}
where $x \in \Xc \subseteq \Rn$ is the state and $u \in \Uc \subseteq \Rm$ is the control input of the system. Both the drift and control vector fields, $f : \Rn \rightarrow \Rn$ and $g : \Rn \rightarrow \Rm$ respectively, are assumed to be Lipschitz continuous. Let the safe state space of the system be encoded as the superlevel set $\Sc$ of a smooth function $h : \Xc \rightarrow R $ as follows,
\vspace{-0.2cm}
\begin{align}\label{safeset}
\mathcal{S} = \{ x \in \Rn \ | \ h(x) \geq 0 \}.
\end{align}
\vspace{-0.5cm}

\noindent \textbf{Definition 1 \cite{cbf}:} \textit{The function $h(x) : \Xc \rightarrow \mathbb{R}$ is defined as a control barrier function (CBF), if $\exists$ an extended class-$\kappa$ function ($\kappa(0) = 0$ and strictly increasing) such that $\forall x \in \Sc$,
\begin{align}\label{cbf}
	\sup\limits_{u \in \Uc } \Big\{ L_fh(x) + L_gh(x)u + \kappa(h(x)) \Big\} \geq 0,
\end{align}
}
where, $L_fh(x)$ and $L_gh(x)$ stand for the Lie derivative of $h(x)$ along the vector fields $f(x)$ and $g(x)$ respectively. 


\noindent \textbf{Theorem \cite{cbf}:} \textit{Given a system defined by (\ref{affine}), with safe set $\Sc \subset \Rn$ defined by (\ref{safeset}), and smooth CBF $h(x) : \Sc \rightarrow \mathbb{R}$ defined by (\ref{cbf}), $\forall$ Lipschitz continuous $u \in \Uc$ that satisfies, $\mathcal{\bar{U}} = \{ u \in \Uc \ | \  L_fh(x) + L_gh(x)u + \kappa(h(x)) \geq 0 \}, \ \forall x \in \Xc$, then the safe set $\Sc$ is forward invariant for the system.
}

CBFs are limited in their nature to systems with relative degree one, i.e., $\delta = 1$, where $\delta \in \mathbb{W}$ \cite{ecbf}. Depending on how one enforces barriers around state(s), relative degree can go above 1. Thus, CBFs cannot be directly applied for such barrier constraints. For $\delta > 1$, an extension of the CBF, called Exponential CBF (ECBF), is used to guarantee forward invariance property of $\Sc$  \cite{ecbf}.

\noindent \textbf{Definition 2 \cite{ecbf}:} \textit{The smooth function $h(x) : \Xc \rightarrow \mathbb{R}$, with relative degree $\delta$, is defined as an exponential control barrier function (ECBF), if $\exists \ \mathcal{K} \in \mathbb{R}^{\delta}$ such that $\forall x \in \Sc$,}
\begin{align}\label{ecbf}
	\sup\limits_{u \in \Uc } \Big\{ L_f^{\delta} h(x) + L_gL_f^{\delta-1}h(x)u + \mathcal{K}^{\top} \mathcal{H} \Big\} \geq 0,
\end{align}
where $\mathcal{H} = [h(x), L_fh(x), L_f^2h(x), ... , L_f^{(\delta-1)} h(x)]^{\top}$ is the vector of Lie derivatives for $h(x)$, and $\mathcal{K} = [k_0, k_1, ..., k_{\delta-1}]$ is the vector of coefficient gains for $\mathcal{H}$. The coeffient gain vector $\mathcal{K}$ can be determined using pole placement technique on the closed-loop matrix $(F-GK)$ determined from $h(x) \geq Ce^{F-GK}\mathcal{H}(x_0) \geq 0$, when $C = [1, 0, 0, ... , 0]^{\top} \in \mathbb{R}^{\delta}$, $h(x_0) \geq 0$ \cite{ecbf}. Forward invariance is also satisfied for ECBFs and we refer the reader to \cite{ecbf} for detailed proofs.

\subsection{Dynamics of 3D Quadrotor}
\begin{figure}[!t]
\centering
\includegraphics[width=0.8\linewidth]{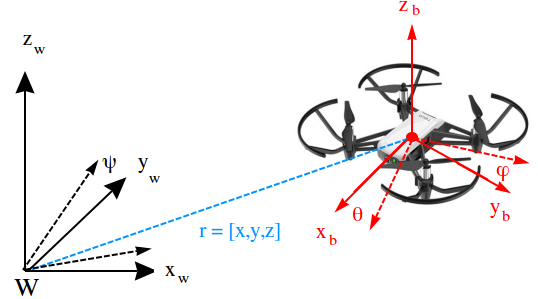}
\vspace{-.4cm}
\caption{World frame $\mathbf{W}$ (black) and body frame $\mathbf{B}$ (red) are shown along with euler angles. The quadrotor's position vector in $\mathbf{W}$ is also marked (blue dashed). DJI Tello Drone is used as base model for illustration.}
\label{fig:quad_frames}
\vspace{-.5cm}
\end{figure}

3D Quadrotor is a dynamical system whose motion evolves in the Lie Group $SE(3)$. Hence, it is described with six degrees of freedom: translational position, $r = [x,y,z]^{\top}$, in the inertial frame and attitude represented by Euler angles, $\eta = [\phi, \theta, \psi]^{\top}$, in the body-fixed frame (see Figure \ref{fig:quad_frames}).
Quadrotor dynamics is well studied in the literature \cite{corkekumar},\cite{schoelligfeed}, hence we only present equations governing dynamics.

The translational acceleration of the quadrotor is \cite{corkekumar},
\begin{align}\label{eq:lin_accel}
	\ddot{r} = gz_w - \mathbf{R}z_{w} \frac{f(t)}{m},
\end{align}
where $z_w = [0, 0, 1]^{\top}$, $m$ is its mass, $g$ is gravity, and $f(t) \in \mathbb{R}$ is the total thrust. $\mathbf{R} \in SO(3)$ is the rotation matrix of body-fixed frame $\mathbf{B}$ and its evolution is given by \cite{corkekumar},
\begin{align}\label{eq:Rdot}
	\mathbf{\dot{R}}(t)
	&=	\mathbf{R}(t) [\Omega(t)]_\times,
\end{align}
where $[\cdot]_\times$ is the overloaded operator for skew-symmetric representation of the angular velocity $\Omega = [p, \ q, \ r]^{\top}$. In the body-fixed frame $\mathbf{B}$, the angular acceleration is given by\cite{corkekumar},
\begin{align}\label{eq:omega_accel}
	\mathbf{I}\dot{\Omega} = \mathbf{\tau} - \Omega \times \mathbf{I}\Omega,
\end{align}
where $\mathbf{\tau} = [\tau_x, \tau_y, \tau_z]^{\top}$ are the moments along each principal axis and $\mathbf{I}$ is the inertia matrix of the quadrotor. The quadrotor system is control affine with its full state as $x = [r, \eta, \dot{r}, \Omega]^{\top}$ and control input $u = [f, \tau_x, \tau_y, \tau_z]^{\top}$.


\section{Nonlinear QP Cascaded Architecture}\label{sec:controller}
We first discuss nonlinear cascaded controller for quadrotors \cite{corkekumar} and present its design. We then discuss details for the barrier-enforced QP modification to the controller. Solving QP online is fast and has been implemented successfully on cruise control \cite{cbf}, bipeds \cite{biped}, and quadrotors \cite{li-swarm-mob}-\cite{li-safe}.

\vspace{-0.2cm}
\subsection{Motivation}
\vspace{-0.1cm}
The cascaded controller is a popular control architecture with practical feasibility and satisfactory performance \cite{quad_dance}, \cite{quad_traj}, \cite{corkekumar}, \cite{schoelligfeed}. The design is intuitive and is commonly used by students, developers, and/or hobbyists. In prior work, CBFs have been successfully deployed on several controllers such as differential flatness and CLF-CBF-QP \cite{3dquad},\cite{diff_flat},\cite{li-safe}. However, augmentation of CBFs to enforce safety in a cascaded control framework, to the best of our knowledge, has not been investigated. By enforcing CBFs in this framework, we aim to provide safety-critical control using a cascaded online QP controller.

\vspace{-0.2cm}
\subsection{Controller Design}
\begin{figure}[!t]
\centering
\includegraphics[width=1\linewidth]{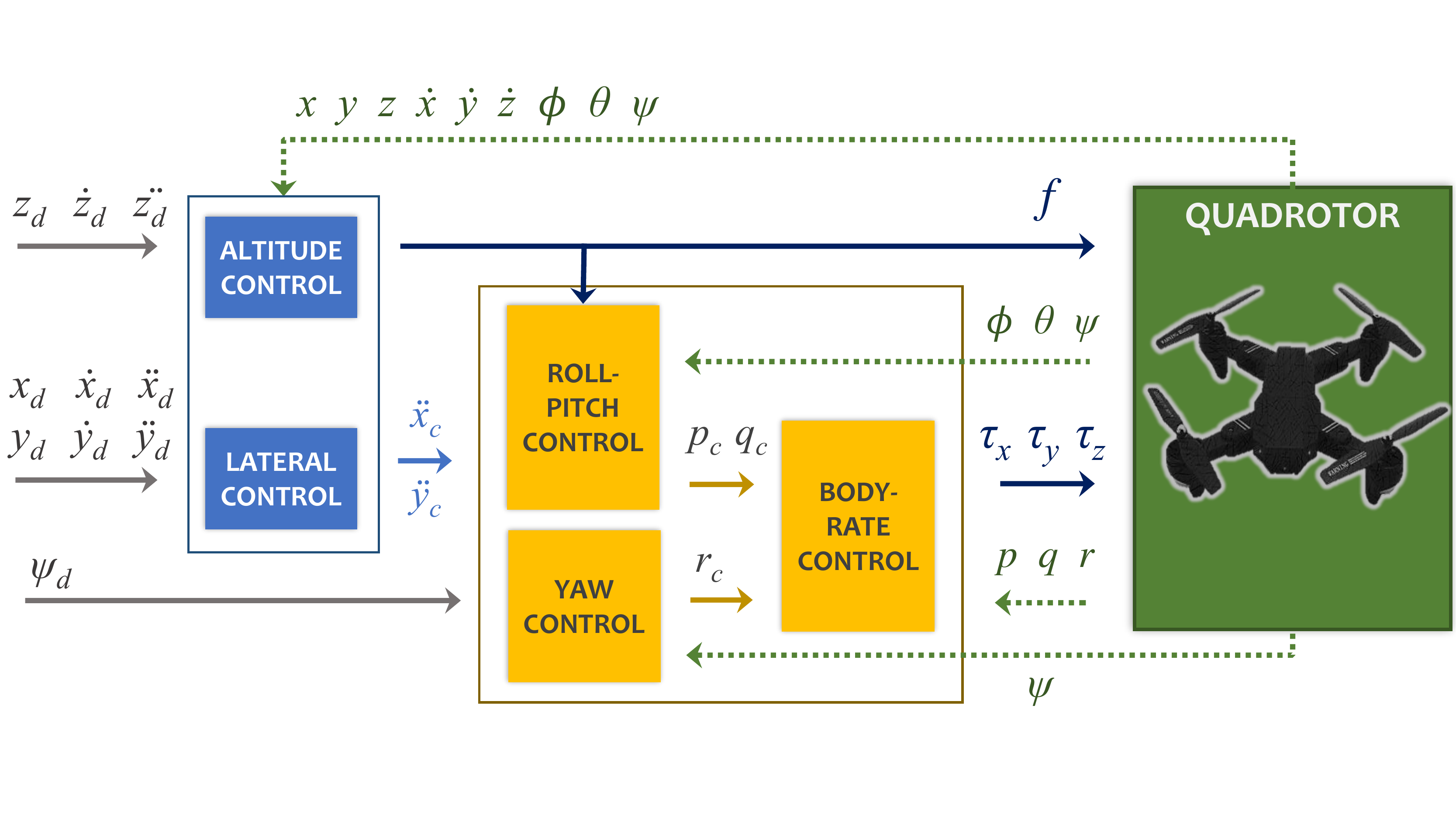}
\vspace{-1.5cm}
\caption{The cascaded controller has a position loop (blue boxed) and attitude loop (yellow boxed). Reference inputs (grey solid) are provided to position and yaw controllers. The altitude controller generates \textbf{desired thrust}. The attitude loop orients roll-pitch and, separately, yaw with the body-rate controller generating \textbf{desired torques}. State measurements (green dotted) to controllers and control inputs (black solid) to quadrotor are shown.}
\vspace{-0.5cm}
\label{fig:quad_architecture}
\end{figure}

\vspace{-0.1cm}
The cascaded terminology is due to the hierarchical approach taken while designing the controllers (see Figure \ref{fig:quad_architecture}). At the highest level is the position controller, which is further decomposed into altitude and lateral controllers. The next level controls the quadrotor's attitude. 
We make the following assumptions as inputs for our controller design:
\begin{itemize}
\item A smooth reference is given: $r_d(t) = [x_d, y_d, z_d]^{\top}$.
\item A yaw reference trajectory is given: $\psi_d(t)$.
\item Observability of states: $x(t) = [r,\eta,\dot{r},\Omega]^{\top}$.
\end{itemize}
Our controller framework is modeled after \cite{schoelligfeed}. The position controller's commanded accelerations are computed as a second-order system: $\ddot{r}_{cmd}(t) = \ddot{r}_d(t) + K_p e_r(t) + K_d \dot{e}_r(t)$, where $e_r(t) = r(t) - r_d(t)$, $K_p$ and $K_d$ are positive definite proportional and derivative gain matrices. Using (\ref{eq:lin_accel}) and altitude commanded acceleration $\ddot{z}_{cmd}$, we get the thrust,
\begin{align}\label{eq:thrust}
f(t) = \frac{m}{R_{33}}(g - \ddot{z}_{cmd} ),
\end{align}

where $R_{33}$ is an entry of $\mathbf{R}$. The attitude controller computes commanded angular velocities using $\mathbf{R}$ and (\ref{eq:Rdot}),
\begin{align}\label{eq:pq_cmd}
\begin{bmatrix}
p_{cmd} \\ q_{cmd}
\end{bmatrix}
&=
\frac{1}{R_{33}}
\begin{bmatrix}
R_{21}	&	-R_{11} \\
R_{22}	&	-R_{12}
\end{bmatrix}
\begin{bmatrix}
\dot{R}_{cmd}^{13} \\
\dot{R}_{cmd}^{23}
\end{bmatrix}
\end{align}
The yaw controller computes $r_{cmd}$ separately since rotations around the quadrotor's $z_B$ axis do not affect the dynamics of roll and pitch. A proportional regulator is used for determining $r_{cmd}$ along $z_B$: $r_{cmd}(t) = k^p_{\psi} (\psi(t) - \psi_d(t))$.

Finally, the body-rate controller in the attitude loop computes $\dot{p}_{cmd}(t), \dot{q}_{cmd}(t), \dot{r}_{cmd}(t)$ using proportional regulators. The moments, $\tau_x, \tau_y, \tau_z$, are then computed using (\ref{eq:omega_accel}).

\subsection{Barrier-enforced QP Cascaded Controller}\label{sub:qp-cascade}

Given the nominal controller $\hat{u} = [f, \tau_x, \tau_y, \tau_z]^{\top}$ developed above, safety barriers are enforced across the cascaded architecture. This ``\textit{modifies}'' the nominal control ensuring the system is always safe. Inspired by cascaded design, we deploy two separate QPs subjected to their own safety constraints, thus achieving independent safety regulation in the altitude and lateral domains. 

The high-level QP responsible for altitude domain is:
\vspace{-0.3cm}
\begin{algorithm}
  High-level QP: \textit{Thrust modification}
	\begin{align}\label{eq:u1-qp-control}
		 F^* = \argmin_{F \in \mathbb{R}} \frac{1}{2} || F - \hat{u}_1 ||^2 \hspace{2cm} \\
		\text{s.t.} \ \ \ \Big\{ L_f^{\delta} h(x) + L_gL_f^{\delta-1}h(x) F + \mathcal{K}^{\top} \mathcal{H} \Big\} \geq 0 \\ \notag
		0 \leq F \leq F_{max}, \hspace{3.6cm}
	\end{align}
\end{algorithm}
\vspace{-0.3cm}
\\
where $\hat{u}_1 = f$ is the nominal thrust computed using (\ref{eq:thrust}), $F^*$ is the computed thrust from QP. Hence, the modified control $F^*$ follows the nominal control $\hat{u}_1$ as closely as possible subject to ensuring safety requirements, thereby, relaxing strict tracking if needed to meet safety requirements.

For ensuring safe control in lateral domain, a low-level QP is designed subject to another set of safety constraints:
\vspace{-0.2cm}
\begin{algorithm}
  Low-level QP: \textit{Torque modification}
	\begin{align}\label{eq:u2-qp-control}
		 M^* = \argmin_{M \in \mathbb{R}^2} \frac{1}{2} || M - \hat{u}_2 ||^2 \hspace{2cm} \\
		\text{s.t.} \ \ \ \Big\{ L_f^{\delta} h(x) + L_gL_f^{\delta-1}h(x) M + \mathcal{K}^{\top} \mathcal{H} \Big\} \geq 0 \\ \notag
		|M| \leq M_{max}, \hspace{4cm}
	\end{align}
\end{algorithm}
\vspace{-0.4cm}
\\
where $\hat{u}_2 = [\tau_x, \tau_y]^{\top}$ are the nominal torques computed and $M^*$ is the resultant torque vector from QP. Construction of Lie derivatives $L_f^{\delta} h(x)$ and $L_gL_f^{\delta-1}h(x)$ for high-level and low-level are discussed below in \ref{sub:high-level-obj} and \ref{sub:low-level-obj} respectively. 

The formulation is thus a cascaded online QP controller that is decoupled in its safety objectives. One layer enforces safety at the high level for altitude domain modifying input $f$ inside the \textit{altitude controller}. The second layer enforces safety at the lower level for lateral domain modifying control inputs $[\tau_x, \tau_y]$ inside the \textit{body-rate controller} (Figure \ref{fig:quad_architecture}).

Unlike \cite{2dquad}, \cite{3dquad} which employ a sequential QP based design, and imposing barriers only at the lower level, we place separate barrier constraints at both levels of the cascaded scheme. Moreover, \cite{3dquad} does not consider input bounds while solving for $F$ and $M$ to find feasible solutions for their QP. 
We have empirically verified in simulation that our QP formulation with actuator bounds found feasible solution.
Quadrotors already employed with cascaded controllers, with very minimal modification, can incorporate safety through our method.

\vspace{-.1cm}
\section{Formulation of Safety Barriers}\label{sec:safety-regions}
\vspace{-.1cm}

To ensure quadrotor's safety, we impose limits on position and velocity states using rectellipsoidal safety regions. 
Inclusion of velocity based constraints explicitly alongside position is imperative as it prevents aggresive braking.

\subsection{Rectellipsoidal Safety Barrier Regions}\label{sub:rectellipse}
\begin{figure}[!t]
\centering
\includegraphics[width=0.8\linewidth]{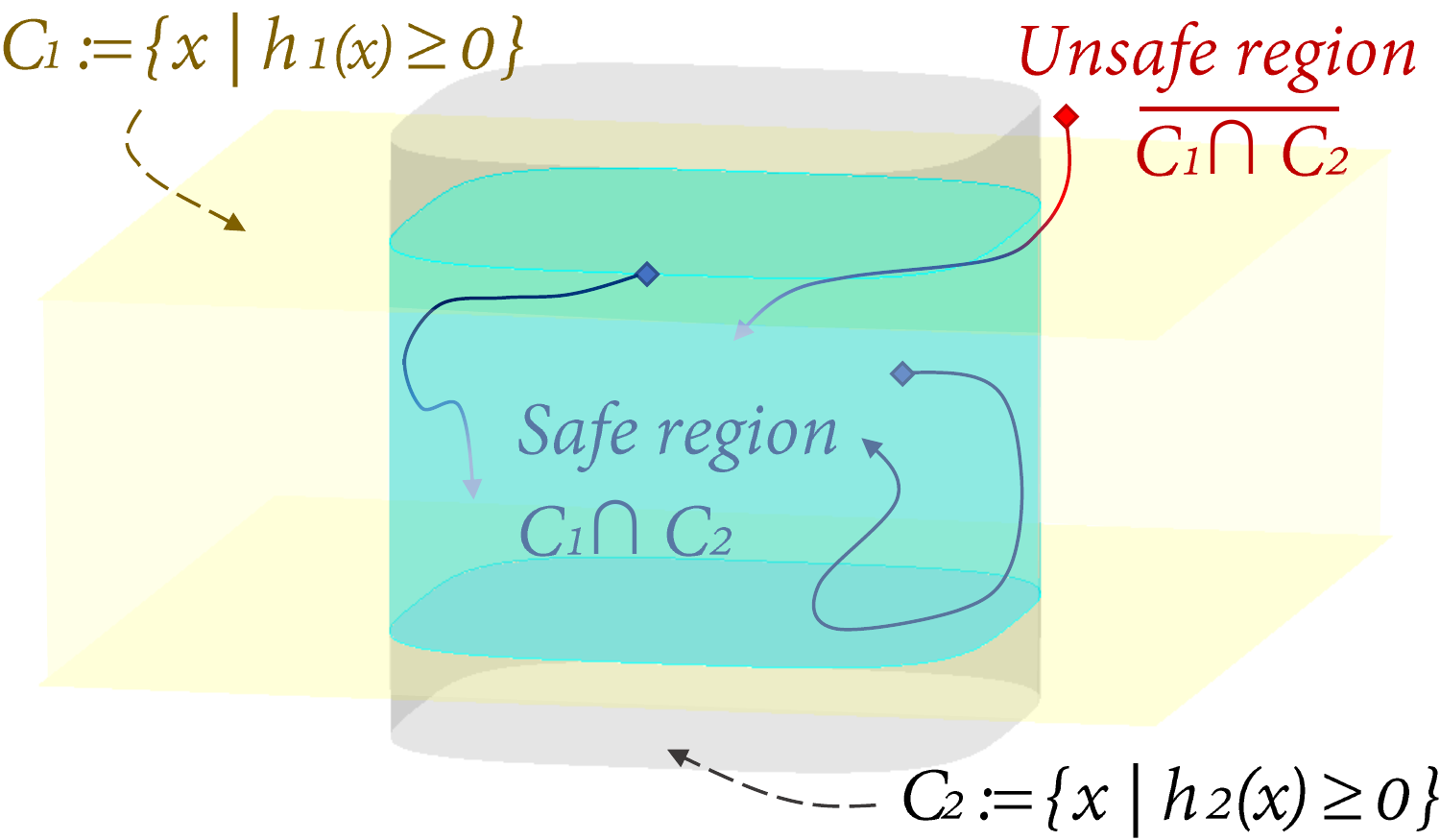}
\vspace{-.5cm}
\caption{Safety barrier regions at altitude and lateral domains ensure forward invariance of states using CBFs. The curves represent state evolution and diamonds represent initial states. Due to imposing separate barriers across the cascaded hierarchy, the intersected volume of safe region $\mathcal{C}_1 \cap \mathcal{C}_2$ results in a richer superlevel set.}
\vspace{-.5cm}
\label{fig:barrier_region}
\end{figure}

The forward invariance property and ellipsoidal model of a safety region is illustrated in Figure \ref{fig:barrier_region}. Inside the safe region, the system's states are allowed to evolve and approach the boundary. Outside the safe region, the control barrier function ensures the system asymptotically approaches the safe region due to CBF constraints (see \cite{cbf} for proof). In our work, the safety barrier region is modeled as,
\begin{align}\label{barrier-region}
h(x_i,...,x_n) = 1 - \Big[ \frac{x_i - c_i}{p_i} \Big]^r + ... + \Big[ \frac{x_n - c_n}{p_n} \Big]^r \geq 0,
\end{align}
where $r$ is the \textit{curve} of the ellipse, $x_i$ is the state of interest, $c_i$ is the ellipse's center, and $p_i$ is the limit enforced on the state. We choose $r = 4$, which is called \textit{rectellipse}, resulting in a richer superlevel set and hence granting greater freedom for determining the safe region (Figure \ref{fig:barrier_region}). Inspired by the work in \cite{li-safe}, where ellipsoidal safe regions (\textit{r = 2}) were used to learn quadrotor dynamics using CBFs in the presence of wind disturbances, we also use a similar safety region for ensuring safety of the quadrotor's state space.

\vspace{-.2cm}
\subsection{High-level Altitude Domain Safety Objective}\label{sub:high-level-obj}
\vspace{-.1cm}
We now look at the high-level safety objective. The overall thrust for the quadrotor is generated by the altitude controller, thereby affecting the quadrotor's altitude position and velocity. In order to enforce limits on the altitude state(s), the following safety barrier region is used,
\begin{align}\label{z-barrier}
h(z) = 1 - \Big[ \frac{z - c_z}{p_z} \Big]^4	\geq 0.
\end{align}
We then compute its Lie derivatives until the control input $u_1 = f(t)$ appears resulting in a relative degree $\delta = 2$. The Lie derivatives are given by,
\begin{align*}
 & \diamond L_fh(z) = \frac{-4(z-c)^3}{p_z^4} \dot{z} \\
 & \diamond L_gL_fh(z) = \frac{4(z-c)^3 R_{33}}{p_z^4 m} \\
 & \ \ \ L_f^2h(z) = \frac{-4(z-c)^3g}{p_z^4} - \frac{12(z-c)^2}{p_z^4}\dot{z},
\end{align*}
where (\ref{eq:lin_accel}) is substituted for $\ddot{z}$. Since relative degree $\delta = 2$, ECBFs are used. The Lie derivatives derived serve as the constraints in (\ref{eq:u1-qp-control}). A single safety region is constructed to handle both position and velocity spaces in altitude domain, with results discussed in Section \ref{sec:simulation}. The barrier function is,
\begin{align}
 \diamond \ h(z,\dot{z}) = 1 - \Big[ \frac{z - c_z}{p_z} \Big]^4 - \Big[ \frac{\dot{z}}{v_z} \Big]^4 \label{zdz-barrier}
\end{align}
Note that for (\ref{zdz-barrier}), $\delta = 1$, hence CBFs are used as opposed to ECBFs for (\ref{z-barrier}). We only present the Lie derivatives for the position space since it has a higher relative degree than the velocity space and the derivation is similar.

\vspace{-.2cm}
\subsection{Low-level Lateral Domain Safety Objective}\label{sub:low-level-obj}
\vspace{-.2cm}
The lower-level safety objective allows enforcing safety limits for movement in the lateral space. The safety limits enforced on lateral positional states $x$ and $y$ are given by,
\begin{align}\label{xy-barrier}
h(x,y) = 1 - \Big[ \frac{x - c_x}{p_x} \Big]^4 - \Big[ \frac{y - c_y}{p_y} \Big]^4 	\geq 0.
\end{align}
Unlike the altitude domain, where the control input $f(t)$ appears directly by computing Lie derivatives, the motion in the lateral plane is affected through the moments $\tau_x$ and $\tau_y$. This involves the effect of roll and pitch to induce this lateral motion. We present the derivation of the dynamic constraints for the low-level QP based controller.

\noindent \textbf{Derivation:} Recall (\ref{eq:pq_cmd}), where angular velocities $p$ and $q$ are related to rotational rates,
\begin{align*}
\begin{bmatrix}
p \\ q
\end{bmatrix}
&=
\frac{1}{R_{33}}
\begin{bmatrix}
R_{21}	&	-R_{11} \\
R_{22}	&	-R_{12}
\end{bmatrix}
\begin{bmatrix}
\dot{R}_{13} \\
\dot{R}_{23}
\end{bmatrix}
=
\frac{1}{R_{33}} W 
\begin{bmatrix}
\dot{R}_{13} \\
\dot{R}_{23}
\end{bmatrix}
\end{align*}
For convenience, we define $W$ as the $2\times2$ matrix of rotational entries and $A \triangleq [p \ \ q]^{\top}$. Rewriting in terms of angular velocities gives,
\begin{align}\label{dotR13R23}
\begin{bmatrix}
\dot{R}_{13} \\
\dot{R}_{23}
\end{bmatrix}
&= R_{33}
W^{-1}
A
=
R_{33} V A \ , \ \ \ \ W^{-1} \triangleq V
\end{align}
Now, computing the time derivative for (\ref{dotR13R23}) results in,
\begin{align}\label{ddotR13R23}
\begin{bmatrix}
\ddot{R}_{13} \\
\ddot{R}_{23}
\end{bmatrix}
&= \dot{R}_{33} V A + R_{33} \dot{V} A + R_{33} V \dot{A} 
\end{align}
Since angular accelerations $\dot{p}$ and $\dot{q}$ are related to inputs $\tau_x$ and $\tau_y$ given in (\ref{eq:omega_accel}), substituting back in (\ref{ddotR13R23}) gives,
\begin{align}\label{JL}
\begin{bmatrix}
\ddot{R}_{13} \\
\ddot{R}_{23}
\end{bmatrix}
&= \dot{R}_{33} V A + R_{33} \dot{V} A + R_{33} V
\begin{bmatrix}
\frac{I_y - I_z}{I_x} qr + \frac{\tau_x}{I_x} \\
\frac{I_z - I_x}{I_y} pr + \frac{\tau_y}{I_y}
\end{bmatrix}			\notag		\\	
&= 
\underbrace{\dot{R}_{33} V A + R_{33} \dot{V} A
+ R_{33}V
\begin{bmatrix}
\frac{I_y - I_z}{I_x} qr \\
\frac{I_z - I_x}{I_y} pr
\end{bmatrix}}_{\mathcal{J}} \notag 
\\
&+
\underbrace{R_{33}V
\begin{bmatrix}
I_x^{-1} & 0 \\
0 & I_y^{-1}
\end{bmatrix}}_{\mathcal{L}}
\begin{bmatrix}
\tau_x \\ \tau_y
\end{bmatrix}	\notag
\\
&= \mathcal{J} + \mathcal{L}
\begin{bmatrix}
\tau_x \\ \tau_y
\end{bmatrix},
\end{align}
where $\mathcal{J}$ and $\mathcal{L}$ are used for simplifying expressions. Since $\ddot{x}$ and $\ddot{y}$ are related to rotational entries $R_{13}$ and $R_{23}$ through (\ref{eq:lin_accel}), we need the fourth time derivative of $x$ and $y$ in order to obtain $\ddot{R}_{13}$ and $\ddot{R}_{23}$, thus finally relating with $\tau_x$ and $\tau_y$. 
\begin{align}
\begin{bmatrix}
\ddot{x} \\ \ddot{y}
\end{bmatrix}
&=
-\frac{f}{m}
\begin{bmatrix}
R_{13} \\ R_{23}
\end{bmatrix}	\hspace{2cm}\text{[using (\ref{eq:lin_accel})]}				\\
\begin{bmatrix}
\dddot{x} \\ \dddot{y}
\end{bmatrix}
&=
-\frac{f}{m} 
R_{33}VA \hspace{2cm} \text{[using (\ref{dotR13R23})] }	\\
\begin{bmatrix}
\ddddot{x} \\ \ddddot{y}
\end{bmatrix}
&=
-\frac{f}{m}\mathcal{J} - \frac{f}{m}\mathcal{L}
\begin{bmatrix}
\tau_x \\ \tau_y
\end{bmatrix} \hspace{0.9cm} \text{[using (\ref{JL})] }
\end{align}
$\hfill\blacksquare$

Thus, time derivatives of $x$ and $y$ relate to control inputs $\tau_x$ and $\tau_y$ with relative degree $\delta = 4$. We next compute the Lie derivatives for lateral safety barrier region (\ref{xy-barrier}),
\begin{align*}
 & \diamond L_fh(x,y) = -4 \eta_3^{\top} \dxy	 \\ 
 & \diamond L_f^2h(x,y) = -4 \eta_3^{\top} \ddxy - 12\eta_2^{\top}\dxysq	\\
 & \diamond L_f^3h(x,y) = -4\eta_3^{\top}\dddxy - 36\eta_2^{\top}\diagdxy\ddxy 
 \\ &  \hspace{1cm}  - 24\eta_1^{\top}\dxycube	\\ \\
 & \diamond L_gL_f^3h(x,y) = \frac{4f}{m}\eta_3^{\top}\mathcal{L} \\
 & \ \ \ L_f^4h(x,y) = \frac{4f}{m}\eta_3^{\top}\mathcal{J} -48\eta_2^{\top}\diagdxy \dddxy \\ 
 & \hspace{1cm} 
 - 36\eta_2^{\top}\ddxysq - 144\eta_1^{\top}\diagdxy\ddxy - 24\eta_0^{\top} \dxyquad,
\end{align*}
where $\eta_i = [ \ (x-c_x)^i/p_x^4 \ , \ (y-c_y)^i/p_y^4 \  \ ]^{\top}, \ \ i \in \{0,1,2,3\}$ and (\ref{JL}) is substituted for $[\ \ddddot{x} , \ \ddddot{y} \ ]^{\top}$. Due to the relative degree being four, ECBFs are once again employed to satisfy the QP constraints in (\ref{eq:u2-qp-control}). For the velocity space of the lateral motion, the following barrier function is used,
\begin{align}\label{dxy-barrier}
h(\dot{x},\dot{y}) = 1 - \Big[ \frac{\dot{x}}{v_x} \Big]^4 - \Big[\frac{\dot{y}}{v_y} \Big]^4 	\geq 0.
\end{align}

Note that, although, higher time derivatives of $x$ and $y$ are present in the Lie derivatives, they are functions of the state ($[r, \eta, \dot{r}, \Omega]^{\top}$). \textit{Numerical differentiation is not required for computing these higher derivatives}. Hence, the issue of having noisy and non-smooth signals that may arise due to differentiation, especially higher order derivatives, is averted.

\section{Simulation Results}\label{sec:simulation}
We present our simulation results using cascaded QP controller from \ref{sub:qp-cascade} and barrier regions constructed at both levels in \ref{sub:high-level-obj} and \ref{sub:low-level-obj}. The simulation was done in MATLAB 2018b with parameters as tabulated in \ref{tab:quad-params} to model the quadrotor. References are generated using sinusoidal curves, $r_d(t) = [a_x \sin(\omega_x t), a_y \sin(\omega_y t), a_z \sin(\omega_z t)]^{\top}$ and $\psi_d(t) = \text{atan2}(y_d,x_d)$. The QP is solved online using MATLAB's built-in \texttt{QUADPROG} solver.
\begin{table}[!b]
\vspace{-.5cm}
\begin{tabular}{lll} \toprule
{Variables} & {Definition} & {Value}\\ \midrule
\text{g} 	& \text{Gravitational acceleration}	& 9.81 \ \text{kg $m/s^2$} \\ 
\text{m} 	& \text{Mass of quadrotor}	& 0.45 \ \text{kg} \\ 
\text{L}	& \text{Distance between two rotors}	& 0.24 \ \text{m}	\\
\text{$I_x$}, \text{$I_y$}	& \text{Inertia about $x_B$-, $y_B$-axis}	& 0.091 \ \text{kg $m^2$}	\\
\text{$I_z$}	& \text{Inertia about $z_B$-axis}	& 0.182 \ \text{kg $m^2$}	\\
\text{$k_f$}	& \text{Motor's thrust constant}	& 0.88 \ \text{m}	\\
\text{$k_w$}	& \text{Motor's torque constant}	& 1.00 \ \text{m}	\\
\text{$f_{min}$}	& \text{Minimum rotor thrust}	& 0.00 \ \text{kg $m/s^2$}	\\
\text{$f_{max}$}	& \text{Maximum rotor thrust}	& 36.00 \ \text{kg $m/s^2$}	\\
\text{$\tau_{min}^{x}$}, \text{$\tau_{min}^{y}$}	& \text{Min. moment about $x_B$, $y_B$-axis}	& -20.0 \text{Nm}	\\
\text{$\tau_{max}^{x}$}, \text{$\tau_{max}^{y}$}	& \text{Max. moment about $x_B$, $y_B$-axis}	&  \ 20.0 \text{Nm}	\\
\bottomrule
\end{tabular}
\vspace{-.4cm}
\caption{Parameters for modeling the dynamics of the quadrotor.}\label{tab:quad-params}
\end{table}

\subsection{High-Level Altitude Domain Safety Behavior}\label{sub:high-sim}
\begin{figure}[!b]
\centering
\includegraphics[width=1.0\linewidth]{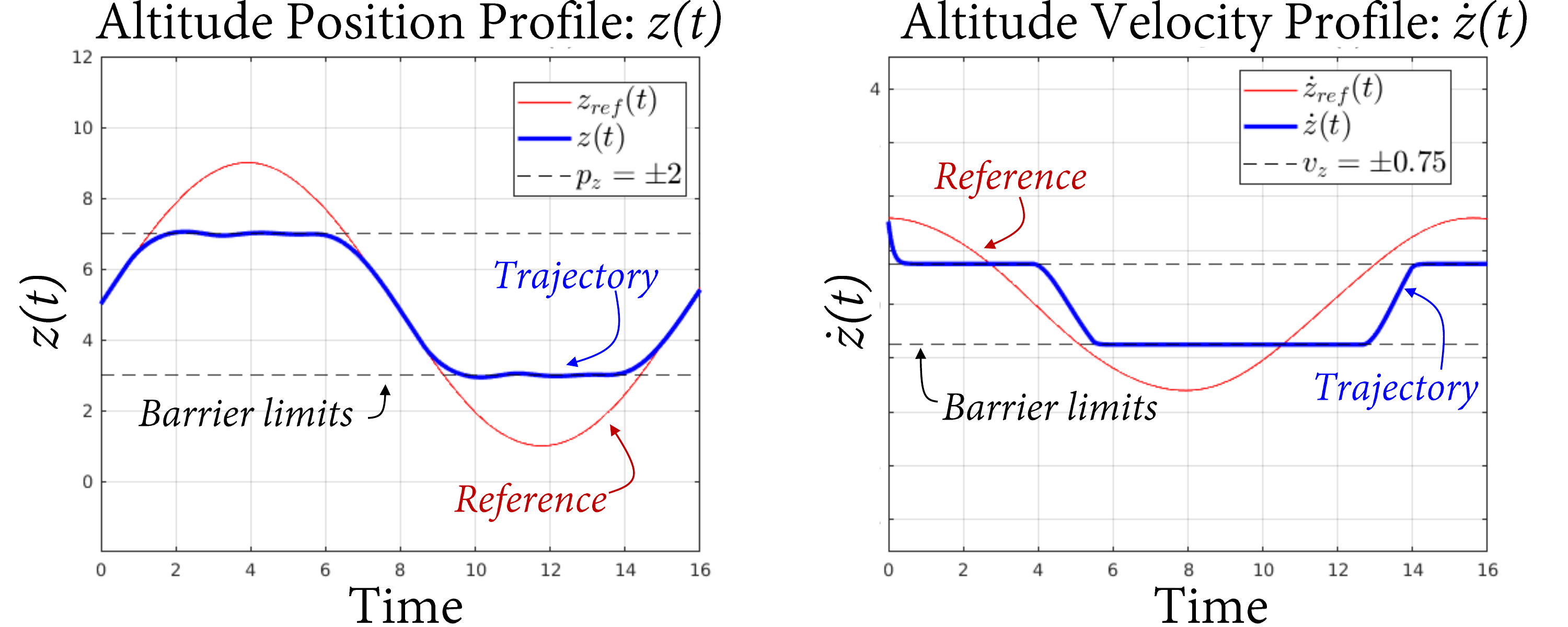}
\vspace{-0.75cm}
\caption{(Left) Position barrier is enforced on state $z$ with a limit of $\pm 2 m$. (Right) Velocity barrier is placed on $\dot{z}$ with $\pm 0.75 m/s$. The rectified trajectory (blue) relaxes tracking the reference trajectory (red) to uphold safety limits (black dashed).}
\label{fig:zdz}
\end{figure}

We impose safety constraints only at the high-level and demonstrate our controller's efficacy in altitude domain. Through the safety barrier regions developed for high-level altitude domain in Section \ref{sub:high-level-obj}, we enforce constraints on altitude position ($z$) and altitude velocity ($\dot{z}$). With barrier limits of $\pm 2 m$ for position and $\pm 0.75 m/s$ for velocity, as shown in Figure \ref{fig:zdz}, trajectory tracking is performed as long as the reference is within the barrier limits. Tracking is relaxed if the reference violates the safety limits.

Note that we subject the quadrotor's altitude reference velocity $\dot{z}_d$ initially to be outside the safe region. The high-level safety objective ensures the quadrotor is first brought into the safe region and contained therein.
\vspace{-.25cm}

\subsection{Low-Level Lateral Domain Safety Behavior}\label{sub:low-sim}
Now, we only impose constraints at the low-level QP responsible for lateral domain. Our cascaded formulation allows easy regulation of quadrotor motion in the lateral domain independent of the high-level constraint objectives since constraints are imposed only at the low-level QP now.

\begin{figure}[!b]
\centering
\includegraphics[width=1.0\linewidth]{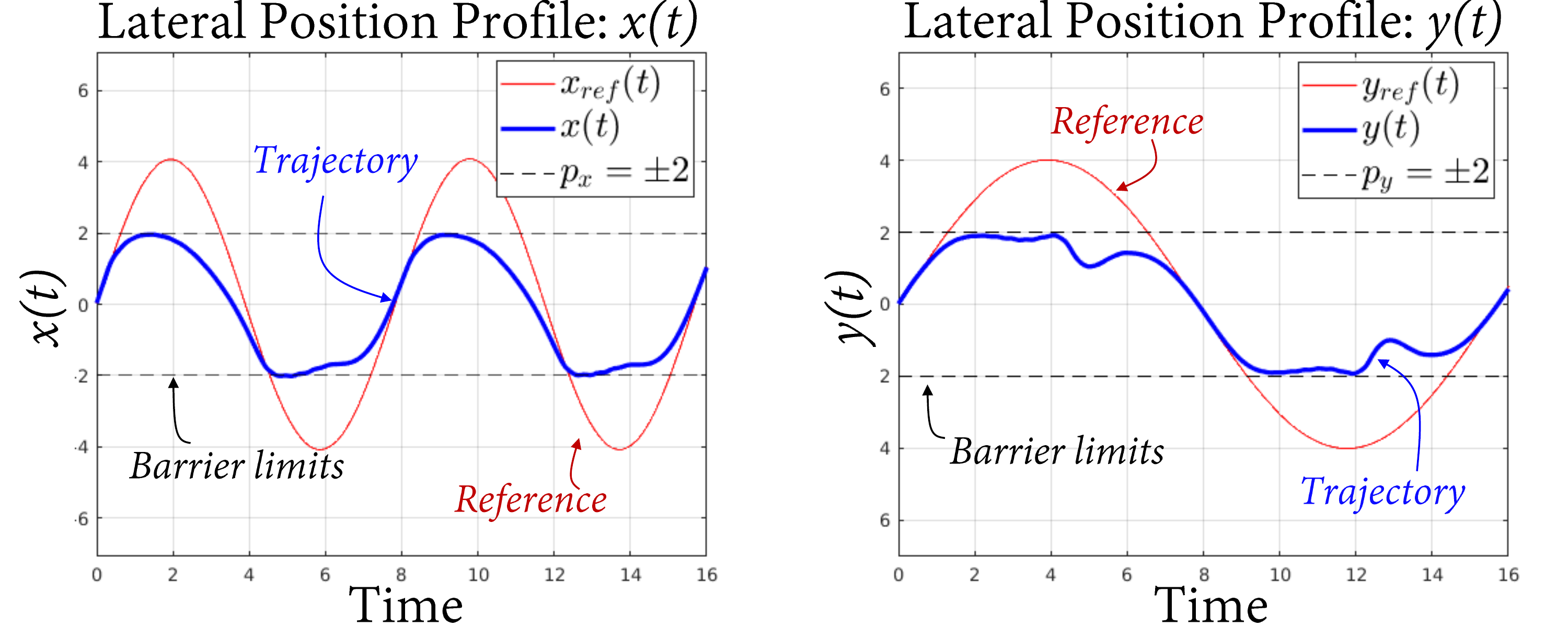}
\vspace{-0.75cm}
\caption{Position barriers are placed on states $x$ (top) and $y$ (bottom) with limits $\pm 2.0$ and $\pm 2.0 m$ respectively. The actual trajectory (blue) compromises the reference trajectory (red) to uphold safety limits (dashed).}
\label{fig:xy}
\end{figure}

\begin{figure}[!b]
\centering
\includegraphics[width=1.0\linewidth]{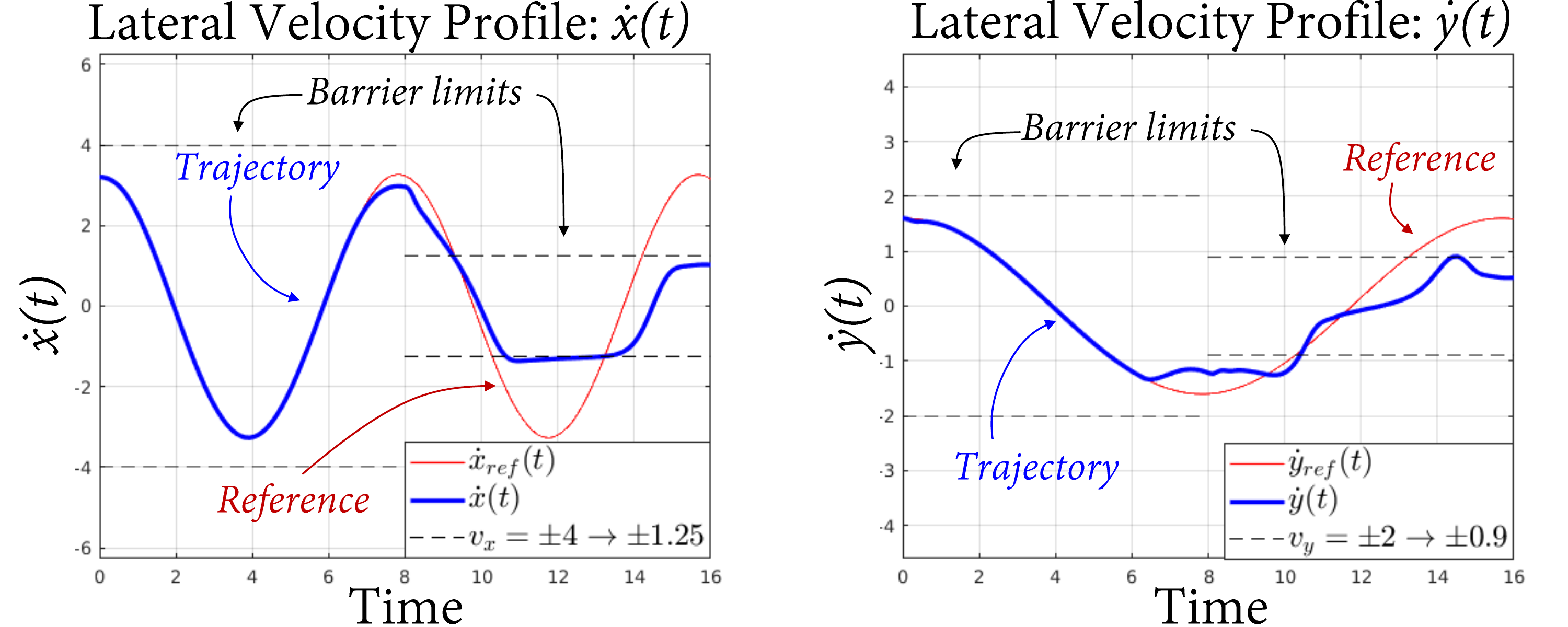}
\vspace{-0.75cm}
\caption{Velocity barriers are enforced on states $\dot{x}$ (top) and $\dot{y}$ (bottom) with initial non-conservative limits of $\pm 4 m/s$ and $\pm 2 m/s$. Barrier limits change mid-flight to more conservative values modifying the controller inputs to respect safety constraints.}
\label{fig:dxy}
\end{figure}

We test our method on both the lateral position ($x,y$) and velocity spaces ($\dot{x},\dot{y}$) and illustrate the results in Figures \ref{fig:xy} and \ref{fig:dxy}. As seen from the two figures, the quadrotor relaxes trajectory tracking when faced with the obligation of upholding safety. This demonstrates that safety barriers are the top priority in regulating the control action.

We also change the velocity barriers mid-way during the flight as shown in Figure \ref{fig:dxy}. For both $\dot{x}$ and $\dot{y}$, initially the barriers were non-conservative values of $\pm 4 m/s$ and $\pm 2 m/s$ respectively. As seen in Figure \ref{fig:dxy}, there is perfect velocity trajectory tracking. The barriers are then restricted to $\pm 1.25 m/s$ and $\pm 0.9 m/s$ for $\dot{x}$ and $\dot{y}$ respectively. The quadrotor reduces its lateral velocities mid-flight in order to respect the barrier constraints.

\begin{figure*}[!t]
\centering
\includegraphics[width=0.85\linewidth]{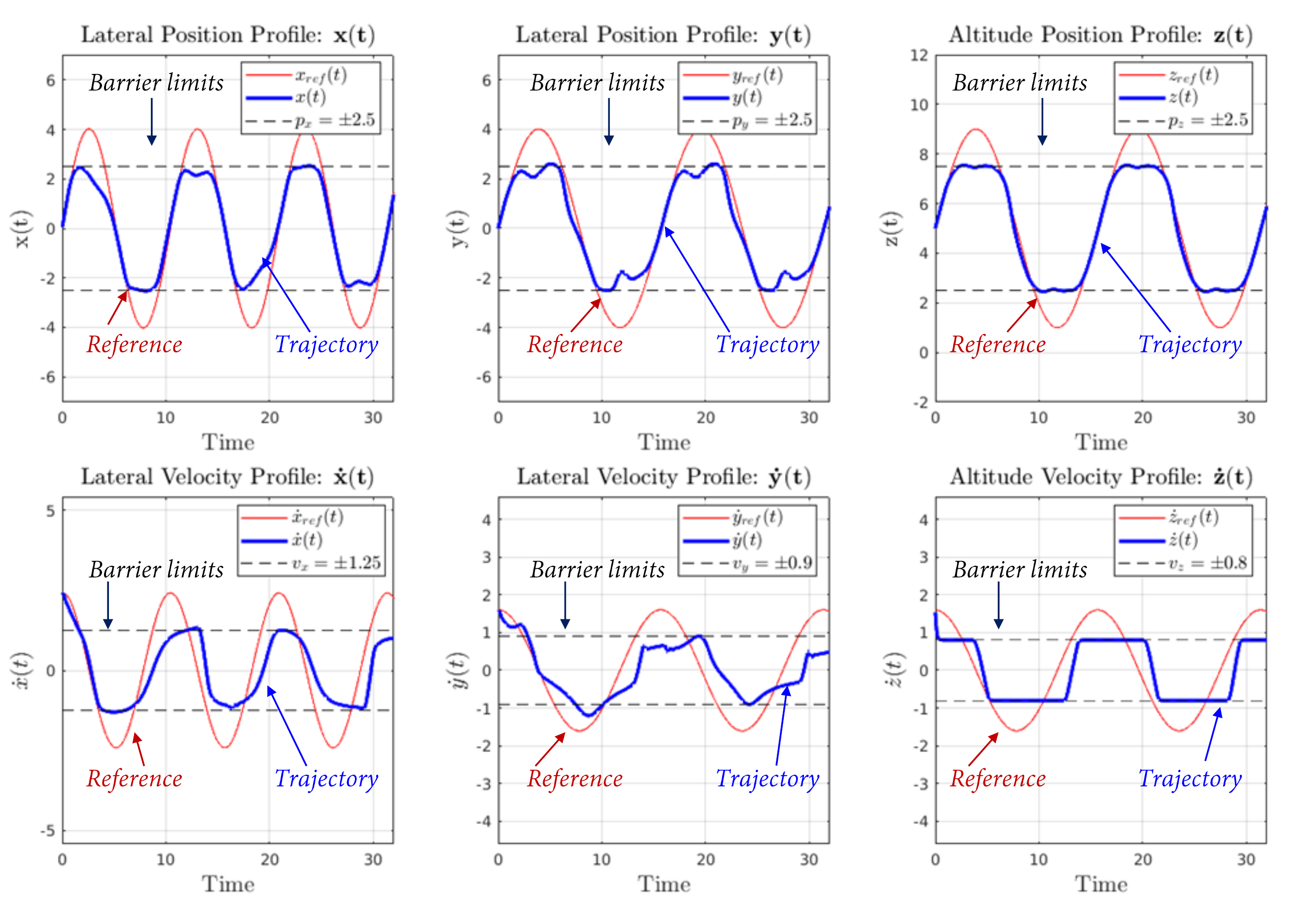}
\vspace{-.5cm}
\caption{Position barriers are placed on $(x,y,z)$ (top) while velocity barriers are placed on $(\dot{x},\dot{y},\dot{z})$ (bottom). The actual trajectory (blue) is the modified flight behavior and the reference trajectory (red) tracking is compromised for respecting safe flight operation given by the barrier limits (black dashed).
}
\label{fig:xy_dxy_zdz}
\vspace{-0.2cm}
\end{figure*}


\begin{figure}[!t]
\centering
\includegraphics[width=1.0\linewidth]{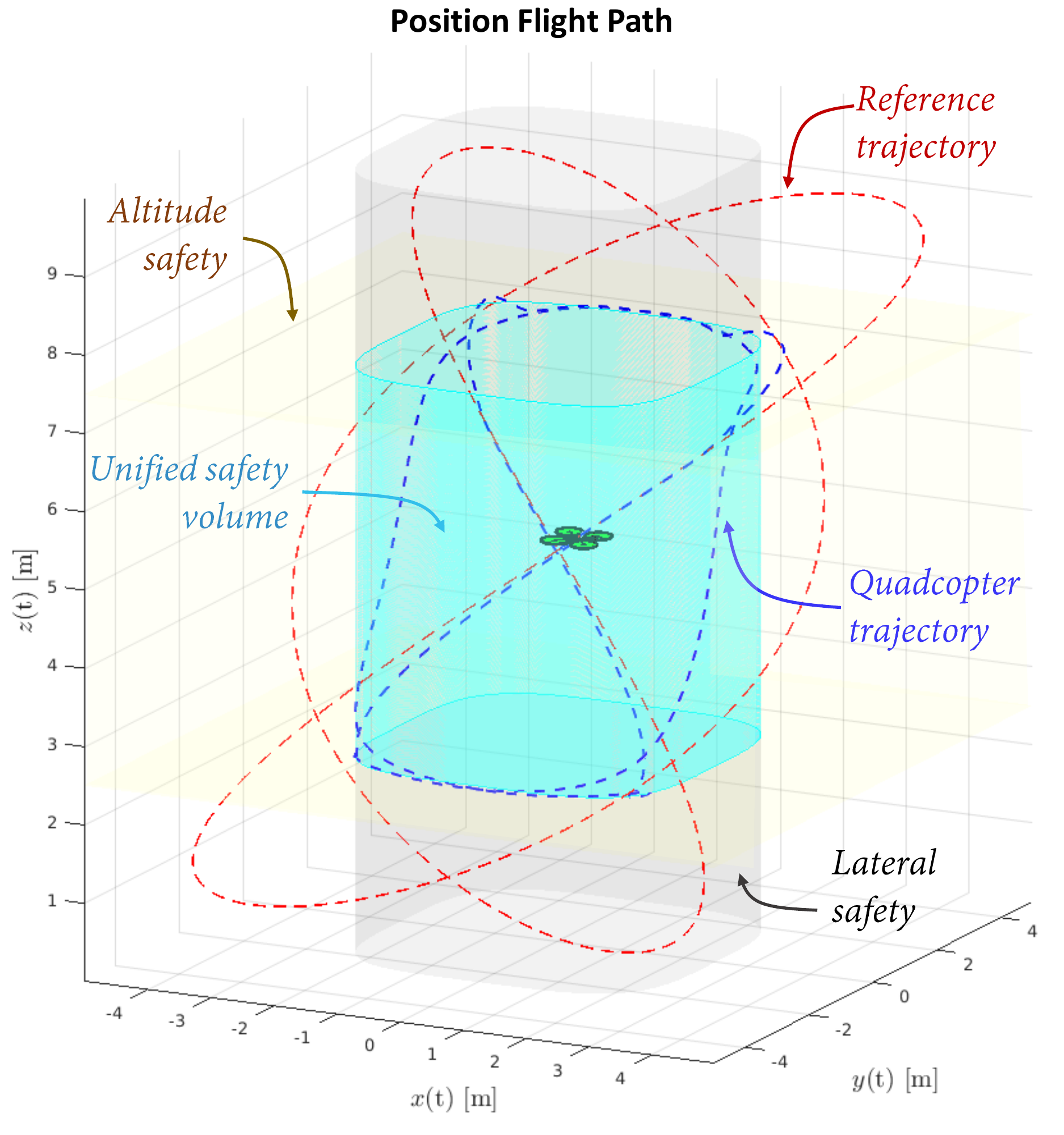}
\vspace{-0.9cm}
\caption{The altitude (yellow) and lateral (gray) safety barrier functions jointly form the unified superellipsoid (cyan). The reference trajectory (dashed red) often violates the safety volume and goes outside. The cascaded QP controller regulating safety at both altitude and lateral domains constrains the quadrotor trajectory (dashed blue) inside the unified region.}
\label{fig:position_flight_path}
\end{figure}

\subsection{Unified Safety Behavior}\label{sub:combined}
In this section, we demonstrate that by applying barriers across the hierarchy, the cascaded QP formulation does not compromise safety in $SE(3)$ and its tangent bundle. Safety is respected in a unified fashion for the quadrotor with each level meeting their safety objectives. The quadrotor is subjected to barrier constraints at both the high-level and low-level domains with safety barrier regions encoded in (\ref{zdz-barrier}) and (\ref{xy-barrier}). By enforcing barriers at both levels, we regulate and ensure safety for the quadrotor's motion in $SE(3)$ domain.

The quadrotor is enforced with different limits for both position $(x,y,z)$ and velocity $(\dot{x},\dot{y},\dot{z})$ states. Moreover, for testing the robustness of meeting the safety objectives at two different levels, quadrotor's initial $\dot{x}$ and $\dot{z}$ velocities are outside their respective safety regions. The trajectory rectification for the different states is illustrated in Figure \ref{fig:xy_dxy_zdz}. As seen in the figure, for each barrier-enforced state, the safety objectives are respected. Even if a particular state is outside the safety region, the constraints ensure the quadrotor asymptotically enter the safety region.

The flight path of the quadrotor is depicted in Figure \ref{fig:position_flight_path}. The safe set is the intersection of the two safe sets, namely, \textit{altitude safety set} and \textit{lateral safety set}. The intersection of these two safe sets results in a richer superlevel safe set, \textit{unified safety set}. The quadrotor's trajectory is constrained inside the \textit{unified safe set} despite the reference trajectory going outside.

\section{Concluding Remarks}\label{sec:conclusion}
In this paper, we demonstrate the augmentation of (exponential) control barrier functions on a nonlinear cascaded control architecture for a quadrotor. We provide separate QP formulations in a cascaded architecture with the high-level safety objective regulating the altitude domain while the low-level safety objective regulating the lateral domain. Despite decoupling the objectives, safety is still preserved in a unified manner for the quadrotor navigation. We demonstrate the effectiveness of our strategy on position and velocity spaces for the quadrotor with both static and dynamic barrier limits.

Despite the effectiveness of our approach, we would like to add some closing remarks on the drawbacks we experienced. Depending on the nature of the barrier region and saturation constraints placed on thrust and moments, there is a possibility for infeasible solutions, thus rendering the QP-based cascaded controller ineffective. We have not found a way to counteract this issue yet. We believe this will be an interesting research direction to investigate further. 
In the future, we would also like to extend the notion by composing several safety barrier regions encapsulating an overall safe volume of space for the quadrotor to navigate.
\vspace{-.1cm}

\section{Acknowledgements}
\vspace{-.1cm}
This work is supported in part by National Science Foundation under grant CISE:S\&AS:1723998.
\vspace{-.1cm}

\bibliographystyle{ieeetr}
\bibliography{root}
\end{document}